\documentclass[aps,prl,twocolumn,groupedaddress,showpacs]{revtex4}

\usepackage{graphicx}

\begin{document}

\title{Saddle-point van Hove singularity and the phase diagram of high-$T_c$ cuprates}

\author{J.G. Storey$^1$, J.L. Tallon$^{1,2}$ and G.V.M. Williams$^1$}

\affiliation{$^1$School of Chemical and Physical Sciences, Victoria
University, P.O. Box 600, Wellington, New Zealand}

\affiliation{$^2$MacDiarmid Institute for Advanced Materials and
Nanotechnology, Industrial Research Ltd., P.O. Box 31310, Lower
Hutt, New Zealand.}

\date{\today}

\begin{abstract}
We examine the generic phase behavior of high-$T_c$ cuprate
superconductors in terms a universal van Hove singularity
in the strongly overdoped region. Using a rigid ARPES-derived
dispersion we solve the BCS gap equation and show that the pairing
interaction or pairing energy cutoff must be a rapidly declining
function of doping. This result is prejudicial to
a phonon-based pairing interaction and more consistent with a
magnetic or magnetically enhanced interaction.
\end{abstract}


\maketitle

The high-$T_c$ cuprates are remarkable in that, despite their varied
atomic structures and consequent variation in bare electronic band
structure\cite{ANDERSEN}, they exhibit universal phase behaviour.
For example, with the exception of La$_{2-x}$Sr$_x$CuO$_4$ which
seems more prone to stripe instabilities, the thermoelectric power
(TEP) is a universal function of hole concentration,
$p$\cite{OBERTELLI}. Moreover, the evolution with doping of both the thermodynamic
properties and the spin susceptibility is likewise
universal\cite{ENTROPYDATA2}. Most importantly, the overall
temperature-doping phase diagram seems to be
universal\cite{TALLON1}. The common features include: the onset of
superconductivity at $p\approx0.05$, the location of so-called
``1/8$^{th}$ anomalies" at $p\approx0.12$, optimal doping at
$p\approx0.16$, critical doping where the pseudogap closes
at $p\approx0.19$, the pseudogap line $T^*(p)$, and possibly the
superconductor/metal transition at $p\approx0.27$. To these we now
wish to add another apparently common feature, namely the presence
of a van Hove singularity (vHs) in the heavily overdoped region\cite{2212VHS}.

The $E(k)$ dispersion for the hole-doped cuprates exhibits a
saddle-point singularity, the so called van Hove singularity, sited
at ($\pi$,0) on the Brillouin zone boundary. Increased hole doping
moves the Fermi energy, $E_F$, down towards the vHs which it
eventually crosses\cite{2212VHS} and where the density of states
(DOS) diverges. Within a BCS picture the transition temperature is
given by\cite{BCSTHEORY}
\begin{equation}
k_BT_c = 1.14\hbar\omega_p \exp(-{1\over{N(E)V}})
\end{equation}
\noindent so that if the vHs crossing occurs within the
superconducting domain the exponential dependence upon the DOS
should result in a local peak in $T_c$, precisely at the vHs. Of
course the cuprates do exhibit a SC phase curve that passes through
a peak at optimal doping\cite{TALLON1} and there have been many
attempts to explain this phase behavior in terms of a vHs crossing
there\cite{BOUVIER,NEWNS,SZOTEK}. Further, the rise in isotope
effect exponent with underdoping was also explored as a consequence
of a proximate vHs\cite{TSUEI}. Finally, underdoped cuprates exhibit
a pseudogap in the normal-state DOS that, amongst other things,
causes a strong suppression of the spin susceptibility, $\chi_s$, at
low temperature\cite{ALLOUL,ENTROPYDATA2}. Some groups sought to
explain this suppression in terms of a nearby vHs\cite{BOUVIER2} where
the Fermi window, for $T>0$, extends to the far side of the vHs
thereby reducing $\chi_s$.

These ideas all failed for various reasons. The value of $\chi_s$
near a vHs never falls more than 10\% at low temperature and so the
model could never account for the nearly full suppression of $\chi_s$
as $T\rightarrow0$\cite{ALLOUL,ENTROPYDATA2}. Further, ARPES revealed that
the vHs at $(\pi,0)$ lay at least 60meV below $E_F$ at
optimal doping and thus could not account directly for the location
of the maximum in $T_c$.

However, we now need to revisit these ideas in view of recent
findings. In the case of La$_{2-x}$Sr$_x$CuO$_4$ the vHs is known to
be crossed in the deeply overdoped region at
$p=x=0.23-0.24$\cite{INO,YOSHIDA}. A similar situation occurs with
Bi$_2$Sr$_2$CuO$_6$\cite{2201VHS} and deeply overdoped
Tl$_2$Ba$_2$CuO$_6$ also lies close to the saddle-point
vHs\cite{PLATE}. We will return to these systems later. They are all
single CuO$_2$ layer compounds. Where there are two CuO$_2$ layers
per unit cell the weak electronic coupling between the layers lifts
the degeneracy of the electronic states in the layers causing
split antibonding and bonding bands. The splitting is maximal near
$(\pi,0)$ and is about 100 meV there. Surprisingly it has recently
been shown by ARPES measurements that for Bi$_2$Sr$_2$CaCu$_2$O$_8$
the antibonding vHs is crossed in the deeply overdoped region
around $p\sim 0.225$\cite{2212VHS}. One has therefore to consider
the hypothesis that the location of a vHs around $p\sim 0.23$ is
general and possibly plays a central role in defining the generic
phase curve $T_c(p)$ (along with the pseudogap at $p_{crit}$). This
is given all the more weight by the fact that both single- and
double-layer cuprates cross the singularity at about the same point
in the phase diagram in spite of the split band in the latter case.

\begin{figure}
\centerline{\includegraphics*[width=70mm]{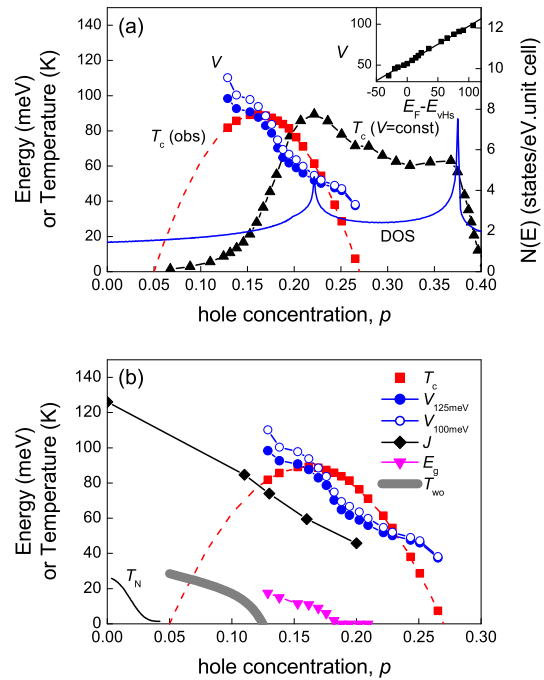}} \caption{\small
(Color online) (a) the doping dependence of the DOS $N(E_F)$ and of $T_c$,
as observed (squares) and as calculated (triangles) assuming a
constant pairing interaction. (b) the doping dependence of the
pairing amplitude, $V$, and of the exchange energy $J$,
the pseudogap energy $E_g$, the NMR wipe-out line $k_BT_{wo}$ and
N\'{e}el line $k_BT_N$.} \label{FIG1}
\end{figure}

We have explored this hypothesis using a rigid ARPES-derived $E(k)$
dispersion for Bi$_2$Sr$_2$CaCu$_2$O$_8$ expressed in terms of a
six-parameter tight-binding fit to the dispersion\cite{2212VHS}. The
doping dependence of the resultant density of states at the Fermi
level, $N(E_F)$, is shown in Fig.~\ref{FIG1}(a). Elsewhere we showed that the
entropy and superfluid density calculated from this dispersion
agrees with the observed magnitude, temperature and
doping dependence of these parameters\cite{STOREY}. Here we ignore
the pseudogap in underdoped samples and this is a feature that would
need to be added in a fuller treatment (though the general
conclusions would remain unchanged).

In the superconducting state we employ a $d$-wave gap of the form
$\Delta_{\textbf{k}}=\frac{1}{2}\Delta_{0}g_{\textbf{k}}$ where
$g_{\textbf{k}}=\cos{k_{x}}-\cos{k_{y}}$. The dispersion in the
presence of the superconducting gap is given by
$E_{\textbf{k}}=\sqrt{\epsilon^{2}_{\textbf{k}}+\Delta^{2}_{\textbf{k}}}$
and $\Delta_{0}(T)$ is determined from the self-consistent
weak-coupling BCS gap equation\cite{BCSGAP}
\begin{equation}
1=\frac{V}{2}\sum_{\textbf{k}}\frac{|g_{\textbf{k}}|^{2}}{E_{\textbf{k}}}\tanh\left(\frac{E_{\textbf{k}}}{2k_{B}T}\right)
\label{BCSGAPEQ}
\end{equation}
For a justification of this approach see Ref.~\cite{STOREY}.
We adopt a pairing potential of the form
$V_{\textbf{kk}^{\prime}}=Vg_{\textbf{k}}g_{\textbf{k}^{\prime}}$.
The summation in Eqn.~\ref{BCSGAPEQ} extends over all states up to a constant energy cut-off $\omega_c$=125meV. We consider two cases: (i) the
pairing amplitude, $V$, is constant, chosen such that $T_{c,max}$ takes the
observed value; and (ii) for each $p$-value $V$ is selected
such that $T_c(p)$ follows the experimentally-observed,
approximately parabolic, phase curve. Fig.~\ref{FIG1}(a) shows $T_c(p)$ plotted
as a function of hole concentration for the two cases.

Turning first to the $T_c(p)$ curve for a constant $V$
(up-triangles), it is evident that if the pairing interaction or
cut-off energy is fixed then the phase curve (a) is more narrow than
that which is observed, (b) maximises at the location of the vHs in
the heavily overdoped region (not at optimal doping) and (c)
exhibits a second peak at the bonding-band vHs. All three
difficulties are averted, and the peak broadened and shifted back to
the observed optimal doping, only if either the pairing interaction
or the cut-off decreases rapidly with doping. This is a robust
result independent of the particular details that follow. We
illustrate this by the second case explored, as follows.

The second $T_c(p)$ phase curve (squares) follows the empirical,
approximately parabolic, phase curve\cite{TALLON1}
\begin{equation}
T_c = T_{c,max} [1 - 82.6 (p-0.16)^2]
\label{TCEQ}
\end{equation}
\noindent Using the gap equation, we have calculated values of $V$ that
reproduce these $T_c$ values. The resultant values of $V$ are
shown by the circles plotted as a function of $p$ in Fig.~\ref{FIG1}(b). They
descend rapidly towards zero with increasing doping. If, alternatively, $V$ is held constant and
the energy cut-off, $\omega_c$, is varied, essentially the same result is
obtained - a rapidly descending value that vanishes near
$p\approx0.3$. Such a strongly dependent interaction would not
usually be associated with the electron-phonon interaction where the
phonon energy scale is only a weak function of doping. Moreover, for
a phonon mechanism involving motion of atoms lying outside of the
CuO$_2$ plane, this function is likely to be non-universal due to
the different bond lengths and substitutionary doping mechanisms in
each high-$T_c$ superconductor. These results are, rather,
indicative of a magnetic, or magnetically enhanced, mechanism.

Further, the pairing pairing potential, $V$, is clearly large and grows with
underdoping towards the magnitude of $J$ the exchange interaction.
For comparison, we plot in Fig.~\ref{FIG1}(b) the magnitude of $J$ determined from
two-magnon Raman scattering\cite{SUGAI}, (where $J$ is taken, as
usual, as 1/3 the frequency of two-magnon scattering peak). The magnitude and doping dependence of $V$ is very similar to that of $J$, suggesting a close relationship between these. Also plotted (open circles) is $V$ when $\omega_c$=100meV. A similar rapid rall is found showing that the choice of $\omega_c$ is not too critical.

This rapid fall in energy scales with doping is also reflected in serveral other energy scales also shown in Fig.~\ref{FIG1}(b). These are the pseudogap energy
scale, $E_g$ and line $T_{wo}/k_B$ where $T_{wo}$ is
the temperature where NMR intensity wipeout effects are observed,
indicating the onset of inhomogeneous spin and charge
distribution\cite{HUNT,SINGER}. $T_{wo}$ is for (Y,Ca)Ba$_2$Cu$_3$O$_{7-\delta}$\cite{SINGER}. It is also in this region that the
4$\times$4 checkerboard structure is observed in scanning tunneling
spectroscopy\cite{KOHSAKA}. These lines all expand out from the
antiferromagnetic phase curve, $T_N(p)$\cite{NISHIDA}, like ripples of remanent
magnetic effects suggesting a common magnetic origin for these.

Here the doping is estimated from the parabolic phase
curve which we know to be approximate only. In fact $V$ is very linear in $E_F-E_{vHs}$, as shown in the inset to Fig.~\ref{FIG1}(a). This suggests that
the overall phase curve $T_c(p)$ is indeed governed by the proximate
vHs combined with a rapidly declining bosonic
energy scale. The value of $\omega_c$ or $V$ need not vanish at
$p\approx0.27$. Eventually the superconducting energy gap will fall
below the pairbreaking scattering rate and $T_c$ will be reduced to
zero\cite{OURWORK3} even if $\omega_c$ or $V$ are not quite zero.

\begin{figure}
\centerline{\includegraphics*[width=70mm]{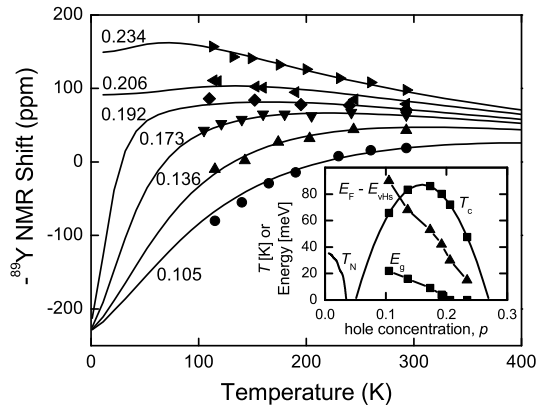}}
\caption{\small The $^{89}$Y Knight shift for
Y$_{0.8}$Ca$_{0.2}$Ba$_2$Cu$_3$O$_{7-\delta}$ at the indicated $p$-values. Solid
curves are the fits using a rigid two-band dispersion, as described
in the text. The inset shows the deduced values of the pseudogap
energy, $E_g$, and of $E_F - E_{vHs}$ showing the vHs is crossed in the deeply overdoped region.}
\label{FIG2}
\end{figure}

If the overall phase diagram in the above-noted single-layer
cuprates and in Bi-2212, is controlled by the proximate vHs then one
might expect a similar vHs crossing in strongly overdoped
Y$_{0.8}$Ca$_{0.2}$Ba$_2$Cu$_3$O$_{7-\delta}$. Here we now present
evidence for this from the $T$-dependent $^{89}$Y Knight shift
measurements reported previously by Williams {\it et
al.}\cite{WILLIAMS2} and shown in Fig.~\ref{FIG2}. The Knight shift
is linearly related to the spin susceptibility, $\chi_s$, as follows
\begin{equation}
^{89}K(T) = a \chi_s(T) + \sigma_c. \label{KNIGHT}
\end{equation}
where $\sigma_c$ is the $T$- and $p$-independent chemical shift. For a Fermi liquid the spin
susceptibility is 
\begin{equation}
\chi_s = 2\mu_B \int {(\partial f/\partial E) N(E)} dE. \label{CHI}
\end{equation}
Here $\mu_B$ is the Bohr magneton and $f(E)$ is the Fermi function.
We adopt our previous model which we used to model the entropy and
superfluid density\cite{STOREY}. Using a tight-binding two-band
fit (provided by the authors of Ref.~\cite{BORISENKO}) to the electronic dispersion for
YBa$_2$Cu$_3$O$_{7-\delta}$ and assuming a rigid band structure with
doping we have calculated $\chi_s$ and fitted to the Knight shift
data shown in Fig.~\ref{FIG2}. As before\cite{STOREY}, we adopt a
Fermi-arc model of the pseudogap and ignore all states in the
integral with energy below the $k$-dependent gap value
$E_g(\theta)$. The fits are shown by the solid curves in the figure
and the deduced values of $E_g$ (=$E_g(\theta =0)$ and $E_F -
E_{vHs}$ are plotted in the inset as a function of doping.

Just like $S^{el}/T$, where $S^{el}$ is the electronic entropy, the
Knight shift is observed to exhibit a fanning out behavior as
$T$ is decreased, with underdoped samples decreasing (due to
the pseudogap) and overdoped samples increasing (due to the
proximate vHs)\cite{STOREY,ENTROPYDATA2}. In fact it has previously
been shown that $\chi_s \approx a_W S^{el}/T$ over a broad range of
$T$ and $p$ for a number of cuprates, with $a_W$ being equal to the
Wilson ratio for nearly-free electrons\cite{ENTROPYDATA2}. With
increasing doping both $\chi_s(p)$ and $S^{el}(p)/T$ should rise
steadily to a peak at the vHs then fall, as has been
observed in La$_{2-x}$Sr$_x$CuO$_4$\cite{ENTROPYDATA2}. The
apparently classical increase in $S^{el}(p)$ of 1 $k_B$ per added
hole\cite{ENTROPYDATA2} would appear simply to arise from the
increased DOS on approaching the vHs. The inset to
Fig.~\ref{FIG2} shows that for
Y$_{0.8}$Ca$_{0.2}$Ba$_2$Cu$_3$O$_{7-\delta}$ the Fermi level will
reach the vHs in the deeply overdoped region, just like the
other cuprates.

We therefore believe this situation to be universal in the cuprates.
It raises the issue as to whether the phase diagram is truly a
universal function of hole concentration, $p$, or whether it
is possibly merely a universal function of the distance away from
the vHs i.e. $E_F - E_{vHs}$. From our point of view this
would be a significant concession as we have long argued for a
universal $T$-$p$ phase diagram\cite{TALLON1,OURWORK3}. But, it is
not straightforward to determine doping levels except in
the case of La$_{2-x}$Sr$_x$CuO$_4$ and
La$_{2-x}$Sr$_x$CaCu$_2$O$_6$ which both follow the universal phase
curve\cite{TALLON1}. To these we can add the fully deoxygenated
material Y$_{1-x}$Ca$_x$Ba$_2$Cu$_3$O$_{6}$ where $p=x/2$, exactly.
Here the universality of the phase diagram seems to be
preserved\cite{TALLON1} and the careful use of bond valence sums
also seems to confirm this for
Y$_{1-x}$Ca$_x$Ba$_2$Cu$_3$O$_{7-\delta}$ when
$\delta<1$\cite{TALLON1}.

Turning to other cuprates, very precise measurements of the cation
content of oxygen-stoichiometric
Tl$_{0.5}$Pb$_{0.5}$Sr$_2$Ca$_{1-x}$Y$_x$Cu$_2$O$_7$,
Tl$_{0.5}$Pb$_{0.5}$Sr$_2$Ca$_2$Cu$_3$O$_9$ and
Tl$_2$Ba$_2$Ca$_2$Cu$_3$O$_{10}$ also confirm the basic phase
diagram with optimal doping at $p\approx0.16$\cite{TALLON1}. For
other cuprates we have usually estimated the doping state either
from the room-temperature thermopower\cite{OBERTELLI} or from the
parabolic phase behavior given by Eqn.~\ref{TCEQ}. HgBa$_2$CuO$_{4+\delta}$ is also found to be consistent with both of these correlations\cite{YAMAMOTO}. But we have recently
found that the overall doping dependence of the thermopower,
$Q(T,p)$, is governed by the approach to the vHs\cite{STOREYTEP2},
with the change in sign of $Q(T,p)$ at low temperature occurring
precisely at the vHs, $p=p_{vHs}$. Thus $Q(T,p)$ is primarily
governed by $(E_F - E_{vHs})$. The thermopower is therefore only a
universal function of $p$ if $(E_F - E_{vHs})$ is a universal
function of $p$. This can be tested by determining the absolute
doping state from the area of the Fermi surface. We will call this
$p_{FS}$.

Consider a few examples. (i) Firstly, it has already been shown that
La$_{2-x}$Sr$_x$CuO$_4$ appears to follow the Luttinger theorem and
$p_{FS}\approx x$ across the phase diagram\cite{YOSHIDA}; (ii) ARPES
measurements on fully-oxygenated YBa$_2$Cu$_3$O$_{6.993}$, which we
have previously identified as having a doping state of
$p$=0.19\cite{OURWORK3}, has a Fermi surface area of 60\% of the
Brillouin zone corresponding to $p$=0.20; and (iii) the Fermi
surface of Tl-2201 has been measured by angular magneto resistance
oscillations\cite{HUSSEY} and for a sample with $T_c$=30K
$p_{FS}$ was found to be 0.24 $\pm$0.02 holes/Cu. Using Eqn.~\ref{TCEQ} its
$T_c$ value implies $p$=0.25. These successes looks very promising.
However, in the case of Bi-2212 the doping state is a little higher
than estimated from thermopower or Eqn.~\ref{TCEQ} and for Bi-2201 it is
seriously higher\cite{MARKIEWICZ}. We thus conclude that it is
likely that the phase diagram is probably a universal function of
$(E_F - E_{vHs})$ rather than of $p$ (though it is often
also a universal function of $p$).

Finally, we note that close to a vHs there should be a tendency to
structural change or local symmetry reduction so as, for example, to
split the vHs and thus to lower the electronic energy. In this
regard it is interesting that most, perhaps all, of the above
mentioned HTS systems tend to be unstable in the heavily overdoped
region. Firstly, the vHs is evidently close to the limits of
overdoping of Bi-2212 and Y,Ca-123. Our experience is that these
systems tend to decompose near these limits. Thus we typically have
to oxygenate at high oxygen pressures using rather low temperatures
($<350^{\circ}$C) to avoid decomposition. For a long time La-214 has
been known to phase separate in this overdoped
region\cite{JORGENSEN} and only by quenching from high temperature
synthesis and reoxygenating (to stoichiometric O=4) at substantially
lower temperatures can this phase separation be
avoided\cite{RADAELLI}. These effects are most noticeable in
polycrystalline samples where there is a large surface to volume
ratio. Single crystals enjoy a metastable state much longer or to
higher temperatures. Oxygenation of Tl-2201 into the heavily
overdoped region results in precipitation of Tl$_2$O$_3$ on the
surface of single crystals or in grain boundaries of polycrystalline
samples. Our attempts to overdope Tl-2201 or
Tl$_{0.5}$Pb$_{0.5}$Sr$_2$CaCu$_2$O$_7$ into this region by Cd
substitution for Tl (generally successful in other circumstances)
failed. And then Bi-2201 undergoes a change in electronic state in
this same region. These various instabilities have always been a
puzzle as has the fact that so many cuprates, as prepared, reside
close to optimal doping. The proximate vHs could be the common
cause.

In summary, we have used a rigid dispersion for
Bi$_2$Sr$_2$CaCu$_2$O$_{8+\delta}$ and computed the phase curve
$T_c(p)$ showing that the only way in which the observed phase curve
may be recovered is if the pairing interaction is a rapidly falling
function of doping. We find that this closely follows the doping
dependence of the exchange interaction, $J$. We also argue that all
HTS cuprates exhibit a van Hove singularity in the deeply overdoped
region and this probably underlies the universal phase diagram
observed in these systems.

We acknowledge funding from the New Zealand Marsden Fund and wish to
thank A. Kaminski for the tight binding fits to the electronic
dispersion for Bi-2212.

\end{document}